\documentclass[aps,pra,twocolumn,superscriptaddress,showpacs,nofootinbib]{revtex4-1}

\usepackage{graphicx}
\usepackage{amsmath}
\usepackage{amssymb}
\usepackage[dvipsnames]{xcolor}
\usepackage[normalem]{ulem}
\usepackage{algorithm}
\usepackage{algorithmicx}
\usepackage{algpseudocode}

\newcommand{\tr}{\mathrm{tr}}

\newcommand{\cB}{\mathcal{B}}

\newcommand{\cH}{\mathcal{H}}

\newcommand{\grad}{\nabla}
\DeclareMathOperator{\proj}{proj}

\begin{document}

\title{Superfast maximum likelihood reconstruction for quantum tomography}

\author{Jiangwei \surname{Shang}}
\email{Current address: Naturwissenschaftlich-Technische Fakult{\"a}t, Universit{\"a}t Siegen, Walter-Flex-Stra{\ss}e 3, 57068 Siegen, Germany.}
\affiliation{Centre for Quantum Technologies, National University of Singapore, Singapore 117543, Singapore}

\author{Zhengyun \surname{Zhang}}
\email{Corresponding email: zhengyun@smart.mit.edu}
\affiliation{BioSyM IRG, Singapore-MIT Alliance for Research and
Technology (SMART) Centre, Singapore 138602, Singapore}

\author{Hui Khoon \surname{Ng}}
\affiliation{Centre for Quantum Technologies, National University of Singapore, Singapore 117543, Singapore}
\affiliation{Yale-NUS College, Singapore 138527, Singapore}
\affiliation{MajuLab, CNRS-UNS-NUS-NTU International Joint Research Unit, %
UMI 3654, Singapore}

\date[]{Posted on the arXiv on March 9, 2017}

\begin{abstract}
Conventional methods for computing maximum-likelihood estimators (MLE) often converge slowly in practical situations, leading to a search for simplifying methods that rely on additional assumptions for their validity. In this work, we provide a fast and reliable algorithm for maximum likelihood reconstruction that avoids this slow convergence. Our method utilizes the state-of-the-art convex optimization scheme---an accelerated projected-gradient method---that allows one to accommodate the quantum nature of the problem in a different way than in the standard methods. We demonstrate the power of our approach by comparing its performance with other algorithms for $n$-qubit state tomography. In particular, an 8-qubit situation that purportedly took weeks of computation time in 2005 can now be completed in under a minute for a single set of data, with far higher accuracy than previously possible. This refutes the common claim that MLE reconstruction is slow, and reduces the need for alternative methods that often come with difficult-to-verify assumptions. In fact, recent methods assuming Gaussian statistics or relying on compressed sensing ideas are demonstrably inapplicable for the situation under consideration here. Our algorithm can be applied to general optimization problems over the quantum state space; the philosophy of projected gradients can further be utilized for optimization contexts with general constraints.
\end{abstract}

\maketitle

\textit{Introduction.---}
Efficient and reliable characterization of properties of a quantum system, e.g., its state or the process it is undergoing, is needed for any quantum information processing task. Such are the goals of quantum tomography \cite{LNP649}, broadly classified into state tomography and process tomography. Process tomography can be recast as state tomography via the Choi-Jamiolkowski isomorphism \cite{QPT1,QPT2}; we hence restrict our attention to state tomography. Tomography is a two-step process: the first is data gathering from measurements of the quantum system; the second is the estimation of the state from the gathered data. This second step is the focus of this article.

A popular estimation strategy is that of the maximum-likelihood estimator (MLE) \cite{MLEreview} from standard statistics. Computing the MLE for quantum tomography is, however, not straightforward due to the constraints imposed by quantum mechanics. While general-purpose convex optimization toolboxes (e.g., CVX \cite{CVX1,CVX2}) are available for small-sized problems, specially adapted MLE algorithms are needed for tackling useful system sizes. Past MLE algorithms \cite{DilML,TeoYS:thesis} incorporate the quantum constraints by going to the \emph{factored space} (see definition later) where the constraints are satisfied by an appropriate parameterization. Gradient methods are then straightforwardly employed in the now-unconstrained factored space. These algorithms can be slow in practice, with an extreme example \cite{8qubit} of an 8-qubit situation purportedly (see Refs.~\cite{CompSens,GaussianNoise,QiLRE,14qubit}) requiring \emph{weeks} of computation time to find the MLE, with bootstrapped error bars (10 MLE reconstructions in all), for the measured data \cite{RoosGuhne}. This triggered a search for alternatives to the MLE strategy \cite{CompSens,GaussianNoise,QiLRE,14qubit,MatrixProd}, specializing to circumstances where certain assumptions about the system apply, permitting simpler and faster reconstruction.

Yet, the MLE approach provides a principled strategy, requiring no extraneous assumptions that may or may not be applicable, and is still one of the most popular methods for experimenters. The MLE gives a justifiable point estimate for the state \cite{Hradil}. It is the natural starting point for quantifying the uncertainty in the estimate: One can bootstrap the measured data \cite{Efron} and quantify the scatter in the MLEs for simulated data; confidence regions can be established starting from the MLE point estimator; credible regions for the actual data are such that  the MLE is the unique state contained in every error region \cite{OER}. It is thus worthwhile to pursue better methods for finding the MLE.

Here, we present a fast algorithm to accurately compute the MLE from tomographic data. The computation of the MLE for a single set of data for the 8-qubit situation mentioned above now takes less than a minute, and can return a far more accurate answer than previous algorithms. The speedup and accuracy originate from two features introduced here: (i) the ``CG-APG" algorithm that combines an accelerated projected-gradient (APG) approach, which overcomes convergence issues of previous methods, with the existing conjugate-gradient (CG) algorithm; (ii) the use of the product structure (if present) of the tomographic situation to speed up each iterative step. The CG-APG algorithm gives faster and more accurate reconstruction whether or not the product structure is present; the product structure, if present, can also speed up previous MLE algorithms. 

\textit{The problem setup}.---
In quantum tomography, $N$ independently and identically prepared copies of the quantum state
are measured one-by-one via a set of measurement outcomes $\{\Pi_k\}_{k=1}^K$, with ${\Pi_k\ge 0}$ $\forall k$ and ${\sum_{k=1}^K\Pi_k={\bf 1}}$. $\{\Pi_k\}_{k=1}^K$  is known as a POVM (positive operator-valued measure) or a POM (probability-operator measurement).  The measured data $D$ consist of a sequence of detection
events $\{e_1,e_2,\ldots, e_N\}$, where $e_\alpha=k$ indicates the click of the detector for outcome $\Pi_k$, for the $\alpha$th copy measured. The likelihood for $D$ given state $\rho$ (the density matrix) is
\begin{equation}
L(D|\rho)=\prod_{k} p_k^{n_k}={\left\{\prod_{k}{\left[\tr(\rho\Pi_k)\right]}^{f_k}\right\}}^N\,,
\end{equation}
where $p_k=\tr(\rho\Pi_k)$ is the probability for outcome $\Pi_k$, $n_k$ is the total number of clicks in detector $k$, and $f_k=n_k/N$ is the relative frequency.

The MLE strategy views the likelihood as a function of $\rho$ for the obtained $D$, and identifies the quantum state $\rho$ that
maximizes $L(D|\rho)$ as the best guess---the MLE. This can be phrased as an optimization problem for the normalized negative log-likelihood, $F(\rho)
\equiv-\frac{1}{N}\log L(D|\rho)$:
\begin{subequations}
\begin{eqnarray}
\underset{\rho\in\cB(\cH)}{\textrm{minimize}}&\quad&F(\rho)
=-\sum_{k=1}^K\nolimits
f_k\log\bigl(\tr(\rho\Pi_k)\bigr)\,,\label{optimization}\\
\text{subject to}&\quad&\rho\ge 0\quad\textrm{and}\quad \tr(\rho)=1\,.\label{eq:constr}
\end{eqnarray}
\end{subequations}
The domain is the space of bounded operators $\cB(\cH)$ on the $d$-dimensional Hilbert space $\cH$.
We refer to \eqref{eq:constr} as the quantum constraints. Any $\rho\in\cB(\cH)$ satisfying \eqref{eq:constr} is a valid state; the convex set of all valid states is the quantum state space. $F$ is convex, and hence has a unique minimum value, on the quantum state space. $F(\rho)$ is differentiable (except on sets of measure zero) with gradient $\grad F(\rho) = -\sum_{k=1}^K\nolimits \Pi_k\,f_k/p_k\equiv -R$, so that $\delta F(\rho)\equiv F(\rho+\delta\rho)-F(\rho)=\tr(\delta \rho\,\grad F)=-\tr(\delta \rho\,R)$ for infinitesimal unconstrained $\delta\rho$.

\textit{The problem of slow convergence}.---
Previous MLE algorithms \cite{DilML,TeoYS:thesis} converge slowly because of the ``by-construction" incorporation of the quantum constraints \eqref{eq:constr}: One writes $\rho=A^\dagger A/\tr(A^\dagger A)$ for $A\in\cB(\cH)$, and performs gradient descent in the \emph{factored space} of unconstrained $A$ operators, for $\widetilde F(A)\equiv F\bigl(\rho=A^\dagger A/\tr(A^\dagger A)\bigr)$. Straightforward algebra yields
\begin{equation}
\delta\widetilde F(A)=-\tr{\left(\delta A\frac{(R-1)A^\dagger}{\tr(A^\dagger A)}+h.c.\right)}.
\end{equation}
to linear order in $\delta A$. $\delta\widetilde F(A)$ is negative---hence walking downhill---for
$\delta A=\epsilon A(R-1)$, for a small $\epsilon$.
This choice of $\delta A$ prescribes a $\rho$-update of the form
\begin{equation}\label{eq:facDescent}
\rho_i\!\rightarrow \!\rho_{i+1}\!=\!\rho_i+\delta\rho_i, \textrm{ with }
\delta\rho_i\!=\!\epsilon[(R-1)\rho_i+\rho_i(R-1)]
\end{equation}
to linear order in $\epsilon$. $\delta\rho_i$ comprises two terms, each with $\rho_i$ as a factor. When the MLE is close to the boundary of the state space---a typical situation when there are limited data (unavoidable in high dimensions) for nearly pure true states---$\rho_i$ eventually gets close to a rank-deficient state with at least one small eigenvalue. Yet, $\rho_i$ has unit trace, so its spectrum is highly asymmetric. $\delta\rho_i$ inherits this asymmetry, leading to a locally ill-conditioned problem and slow convergence.

\begin{figure}[h]
\includegraphics[trim=3mm 3mm 2mm 3mm, clip, width=\columnwidth,height=6cm]{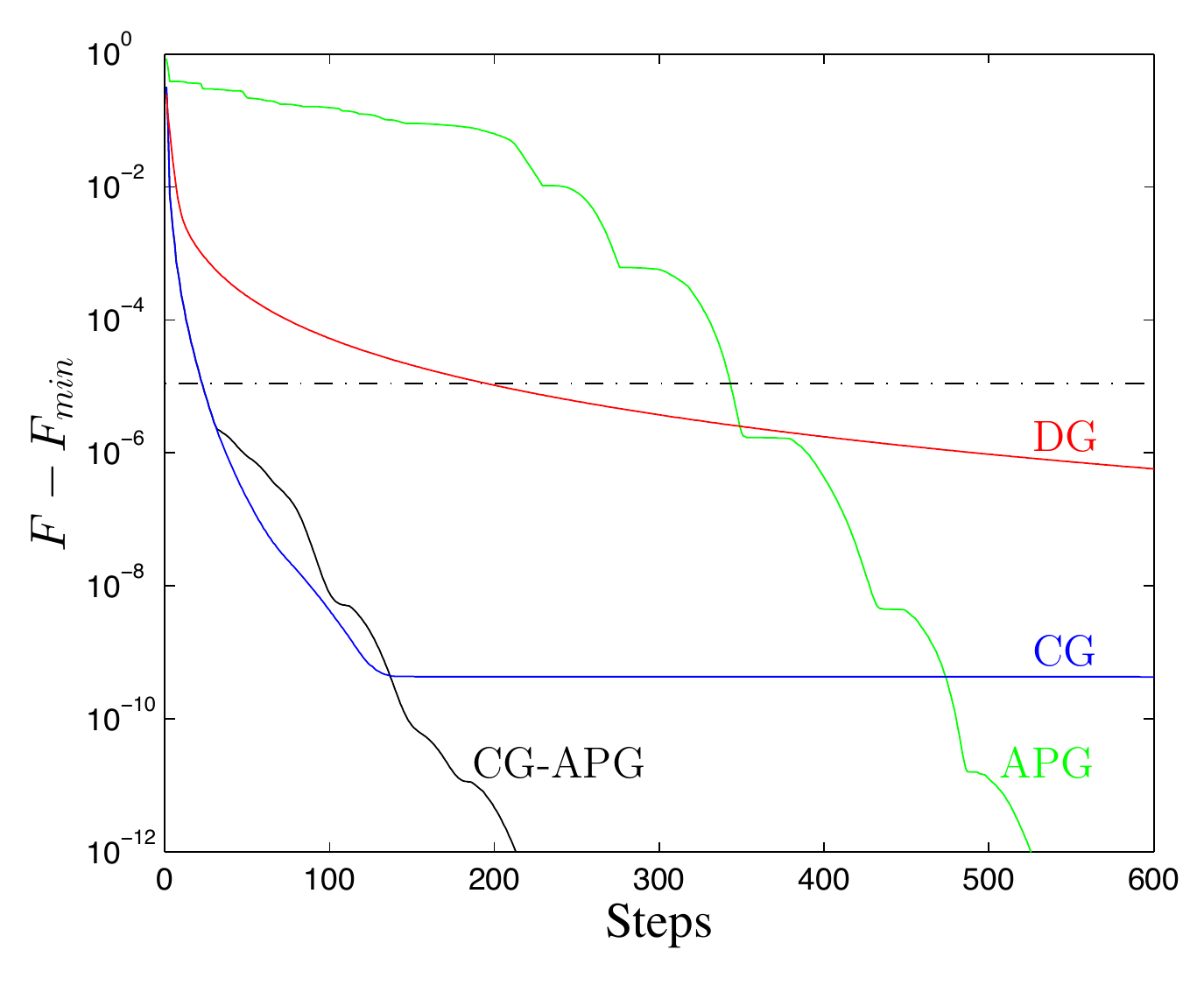}
\caption{\label{fig:8qubit}%
The deviation $F-F_{\min}=-\frac{1}{N}\log(L/L_{\max})$ at each iterative step for different algorithms, for the experimental data of \cite{8qubit}. $F_{\min}$ is the smallest $F$ value attained among the algorithms (reached by APG and CG-APG when run until further progress is hindered by machine precision); $L_{\max}$ is the corresponding likelihood value. Here, $N=3^8\times 100$, for 100 copies for each of the $3^8$ settings of the 8-qubit product-Pauli POM. The dash-dotted line indicates the $F$ value obtained in \cite{8qubit} with the DG algorithm.
}
\end{figure}

To illustrate, consider the situation of \cite{8qubit}: tomography of a (target) $8$-qubit $W$-state via product-Pauli measurements. Figure \ref{fig:8qubit} shows the trajectories taken by different algorithms from the maximally mixed state to the MLE---the minimum of $F$---for the experimental data of \cite{8qubit}. The red and blue lines are for commonly used MLE methods: the diluted direct-gradient (DG) algorithm \cite{DilML} and the CG algorithm with step-size optimization via line search \cite{TeoYS:thesis}. Both algorithms walk in the factored space, with DG performing straightforward descent according to Eq.~\eqref{eq:facDescent}, while CG follows the conjugate-gradient direction.
The DG and CG iterations initially decrease $F$ quickly, but the advances soon stall, with $F$ stagnating at values significantly larger than attainable by APG and CG-APG (explained below). On average the CG-APG and DG algorithms take comparable time per iterative step; see Appendix~\ref{App0} for Fig.~\ref{fig:8qubit} plotted against time rather than steps.

Note that the methods of Refs.~\cite{GaussianNoise, QiLRE,14qubit} are inapplicable here. Those methods assume Gaussian statistics, valid only when every measurement outcome receives many clicks. For the 8-qubit dataset above, 82\% of the outcomes had \emph{zero} counts. This is typical of high-dimensional experiments with limited data. Near-zero counts are also unavoidable for near-rank-deficient states. The compressed-sensing approach \cite{CompSens}, which assumes a low-rank true state, is also a poor choice; see Appendix~\ref{R8qb}.

\textit{The CG-APG algorithm.}---
The slowdown in convergence for DG and CG puts a severe limit on the accuracy of the MLE reconstruction: The analysis of \cite{8qubit} stopped---after a long wait \cite{RoosGuhne}---at a state with likelihood $L\simeq0.1\%L_{\max}$. That was sufficient for \cite{8qubit} to show the establishment of entanglement, but is hardly useful for further MLE analysis. The slowdown in DG and CG can be avoided by walking in the $\rho$-space. There, $F(\rho)$ has gradient $-R$ which, unlike that of $\widetilde F(A)$, is not proportional to $\rho$. Walking in the $\rho$-space, however, does not ensure satisfaction of constraints \eqref{eq:constr}. They are instead enforced by projecting $\rho$ back into the quantum state space after each unconstrained gradient step. This is an example of ``projected-gradient" methods in numerical optimization \cite{Goldstein,Levitin,Bruck,Passty,Nesterov}.

In steepest-descent methods, the local condition number of the merit function [$F(\rho)$ or $\widetilde F(A)$] affects convergence. The condition number measures the asymmetry of the response of the function to changes in the input along different directions. Poor conditioning (i.e., more asymmetric) leads to a steepest-descent direction that oscillates. One smooths out the approach to the minimum by giving each step some ``momentum" from the previous step. CG implements this for quadratic merit functions; for projected gradients, accelerated gradient schemes \cite{FISTA} are the state of the art. Coupled with adaptive restart \cite{AdaptRes}, the APG method indirectly probes the local condition number by gradually increasing the momentum preserved (controlled by $\theta$ in the algorithm below), and resetting ($\theta=1$) whenever the momentum causes the current step to point too far from the steepest-descent direction. The APG algorithm of Refs.~\cite{FISTA,TFOCS,AdaptRes}, in $\rho$-space, thus proceeds as follows:

\begin{algorithm}[H]
\caption{\textbf{APG with adaptive restart}}
\begin{algorithmic}
\State Given $\rho_0$, $0<\beta<1$, and $t_1>0$.
\State Initialize $\varrho_0=\rho_0$, $\theta_0=1$.
\vspace*{0.15cm}
\For {$i = 1,\cdots,$}
\State Set $t_i\!=\!t_{i-1}$, $\rho_i \!= \!\proj(\varrho_{i-1}-t_i\nabla F(\varrho_{i-1}))$, $\delta_i\!=\!\rho_i-\varrho_{i-1}$.
\vspace*{-0.15cm}
\State (Choose step size via backtracking)
\While {$F(\rho_i) > F(\varrho_{i-1}) + \left\langle\nabla
  F(\varrho_{i-1}),\delta_i\right\rangle +
  \tfrac1{2t_i}||\delta_i||_F^2$}
\State Set $t_i=\beta t_i$.
\State Update $\rho_i = \proj(\varrho_{i-1}-t_i\nabla
F(\varrho_{i-1}))$, $\delta_i=\rho_i-\varrho_{i-1}$.
\EndWhile
\vspace*{0.15cm}
\State Set $\hat\delta_i=\rho_i-\rho_{i-1}$; Termination criterion.
\vspace*{0.15cm}
\If {$\langle\delta_i,\hat\delta_i\rangle <0$} \hspace*{0.1cm}(Restart)
\State $\rho_i=\rho_{i-1}$, $\varrho_i=\rho_{i-1}$, $\theta_i=1$;
\vspace*{0.15cm}
\Else \hspace*{0.1cm}(Accelerate)
\State Set $\theta_i=\tfrac{1}{2}{\left(1+\sqrt{1+4\theta_{i-1}^2}\right)}$, $\varrho_i=\rho_i+\hat\delta_i\frac{\theta_{i-1}-1}{\theta_i}$\,.
\EndIf
\EndFor
\end{algorithmic}
\end{algorithm}

\noindent $\proj(\cdot)$ projects the Hermitian argument to the nearest---in Euclidean distance---state satisfying \eqref{eq:constr} \cite{GaussianNoise}. One can modify the backtracking portion of the algorithm for better performance; see Sec.~3 in Appendix~\ref{AppA}. The \textsl{MATLAB} code for our APG and CG-APG algorithms, with accompanying documentation, is available online \cite{qMLE}.

Applying the APG algorithm to the 8-qubit example, one finds fast convergence to the MLE (see Fig.~\ref{fig:8qubit}) once the walk brings us sufficiently close; no slowdown as seen in DG and CG is observed. APG with adaptive restart exhibits linear (on a log-scale) convergence in areas of strong convexity \cite{AdaptRes} sufficiently close to the minimum.
The staircase pattern is expected in adaptive restart APG algorithms \cite{AdaptRes}: Flat regions occur after a reset, giving way to steep regions when the momentum is built up again. These undulations do not affect the overall rate of convergence.

Far from the minimum, APG can descend slowly, as is visible in Fig.~\ref{fig:8qubit}. CG descent in the factored space, on the other hand, is rapid in this initial phase. Similar behavior is observed for other states (see a representative example in Appendix~\ref{AppB}), although the initial slow APG phase is usually markedly shorter than in the $W$-state example here. Thus, a practical strategy is to start with CG in the factored space to capitalize on its initial rapid descent, and switch over to APG in the $\rho$-space when the fast convergence of APG sets in, \emph{provided} one can determine the switchover point cheaply.

Both APG and CG use a local quadratic approximation at each step, the accuracy of which relies on the local curvature, measured by the Hessian of the merit function. The advance is quick if the Hessian changes slowly from step to step so that prior-step information provides good guidance for the next step. Empirically, for nearly pure true states, we observe that the Hessian of $F(\rho)$ changes a lot initially in APG but settles down close to the MLE. This is because the APG trajectory quickly comes close to the state-space boundary, so that some $p_k$ values, which occur in the Hessian of $F(\rho)$ as $\sim\!\! f_k/p_k^2\equiv h_k$, can be small and unchecked by the $f_k$ values away from the MLE. In contrast, the Hessian of $\widetilde F(A)$ relevant for CG is initially slowly changing, but starts fluctuating closer to the MLE, likely due to the ill-conditioning discussed previously. The proposal is hence to start with CG in the factored space, detect when the Hessian of $F(\rho)$ settles down, at which point one switches over to APG in the $\rho$-space for rapid convergence to the minimum. The Hessian is, however, expensive to compute; one can instead get a good gauge by monitoring the $h_k$ values, cheaply computable from the $p_k$s already used in the algorithm; see Sec.~4 in Appendix~\ref{AppA}. This then is finally our CG-APG algorithm, with a superfast approach to the MLE that outperforms all other algorithms; see Fig.~\ref{fig:8qubit}.

\textit{Exploiting the product structure.}---
Part of the speedup in the 8-qubit example stems from exploiting the product structure of the situation. For the four algorithms compared, one of the most expensive parts of the computation is the evaluation of the probabilities $\{p_k=\tr(\rho\Pi_k)\}$ needed in $F$ and $\grad F=-R$, for $\rho$ at each iterative step. For a $d$-dimensional system and $K$ POM outcomes, the computational cost of evaluating $\{p_k\}$ is $O(Kd^2)$ [there are $K$ probabilities, each requiring $O(d^2)$ operations for the trace of a product of two $d\times d$ matrices]. For the 8-qubit example, $d=2^8=256$, and the POM has $K=6^8=1679616$ outcomes.

The computational cost can be greatly reduced if one has a product structure: The system comprises $n$ registers, and the POM is a product of individual POMs on each register. For simplicity, we assume the $n$ registers each have dimension $d_r$, and the POM on each register is the same, written as $\{\pi_k\}_{k=1}^{K_r}$. The $n$-register POM outcome is $\Pi_{\vec k}=\pi_{k_1}\otimes\pi_{k_2}\otimes\ldots\otimes\pi_{k_n}$, with $\vec k\equiv (k_1,k_2,\ldots ,k_n)$ and $k_a=1,\ldots,K_r$. The generalization to non-identical registers and POMs is obvious. Then, $d=d_r^n$ and $K=K_r^n$. Exploiting this product structure reduces the cost of evaluating the probabilities from $O(K_r^n d_r^{2n})$ to $O(K_r^{n+1})$ (for $K_r>d_r^2$). For $n$ qubits with product-Pauli measurements ($d_r=2$, $K_r=6$), this is a huge reduction from $\sim\!6^n4^n$ to $\sim\!6^{n+1}$.

The computational savings come from re-using parts of the calculation. Let $\rho_{n-1}^{(k_n)}\equiv \mathrm{tr}_n(\rho\pi_{k_n})$, the partial trace on the $n$th register, for a given $k_n$. This same $\rho_{n-1}^{(k_n)}$ can be used to evaluate $\rho_{n-2}^{(k_{n-1},k_n)}\equiv \tr_{n-1}{\left(\rho_{n-1}^{(k_n)}\pi_{k_{n-1}}\right)}$ for any $k_{n-1}$. One does this repeatedly, partial-tracing out the last register each time, until one arrives at $p_{\vec k}=\rho_0^{(k_1,k_2,\ldots,k_n)}$.
At each stage, evaluating $\rho_{\ell-1}^{(k_{\ell},\ldots,k_n)}$ from $\rho_{\ell}^{(k_{\ell+1},\ldots,k_n)}$ involves computing the trace of $\pi_{k_\ell}$ with submatrices of $\rho_{\ell}^{(k_{\ell+1},\ldots,k_n)}$. Specifically,\begin{equation}
\rho_{\ell}^{(k_{\ell+1},\ldots,k_n)}=\sum_{\vec i^{(\ell-1)},\vec j^{(\ell-1)}}|\vec i^{(\ell-1)}\rangle\langle \vec j^{(\ell-1)}|\otimes\rho_{\vec i^{(\ell-1)};\vec j^{(\ell-1)}},
\end{equation}
where $\vec i^{(\ell-1)}\equiv (i_1,i_2,\ldots, i_{\ell-1})$ with $i_a=1,\ldots, d_r$ (similarly for $\vec j^{(\ell-1)}$), $\rho_{\vec i^{(\ell-1)};\vec j^{(\ell-1)}}$ is a $d_r\times d_r$ submatrix, and $\rho_{\ell}^{(k_{\ell+1},\ldots,k_n)}$ is a $(d_r)^{\ell-1}\times (d_r)^{\ell-1}$ array of these submatrices. Getting $\rho_{\ell-1}^{(k_{\ell},\ldots,k_n)}$ from $\rho_{\ell}^{(k_{\ell+1},\ldots,k_n)}$ requires replacing each submatrix in $\rho_{\ell}^{(k_{\ell+1},\ldots,k_n)}$ by the number $\tr(\rho_{\vec i^{(\ell-1)};\vec j^{(\ell-1)}}\pi_{k_\ell})$, which takes $O(d_r^2)$ computations. Since each $\rho_{\ell}^{(k_{\ell+1},\ldots,k_n)}$ need only be computed once for all subsequent $k_{j\leq\ell}$, simple counting (see Appendix~\ref{AppC}) yields a total cost of $O(K_r^{n+1})$ (for $K_r>d_r^2$) to evaluate the full set of probabilities. Constructing $R$ for given $\{p_k\}$ also requires $O(Kd^2)$ operations; the same idea of register-by-register evaluation applies for a speedier computation.

\begin{figure}
\includegraphics[trim=3.5mm 3.5mm 4mm 5mm, clip, width=\columnwidth, height=6.3cm]{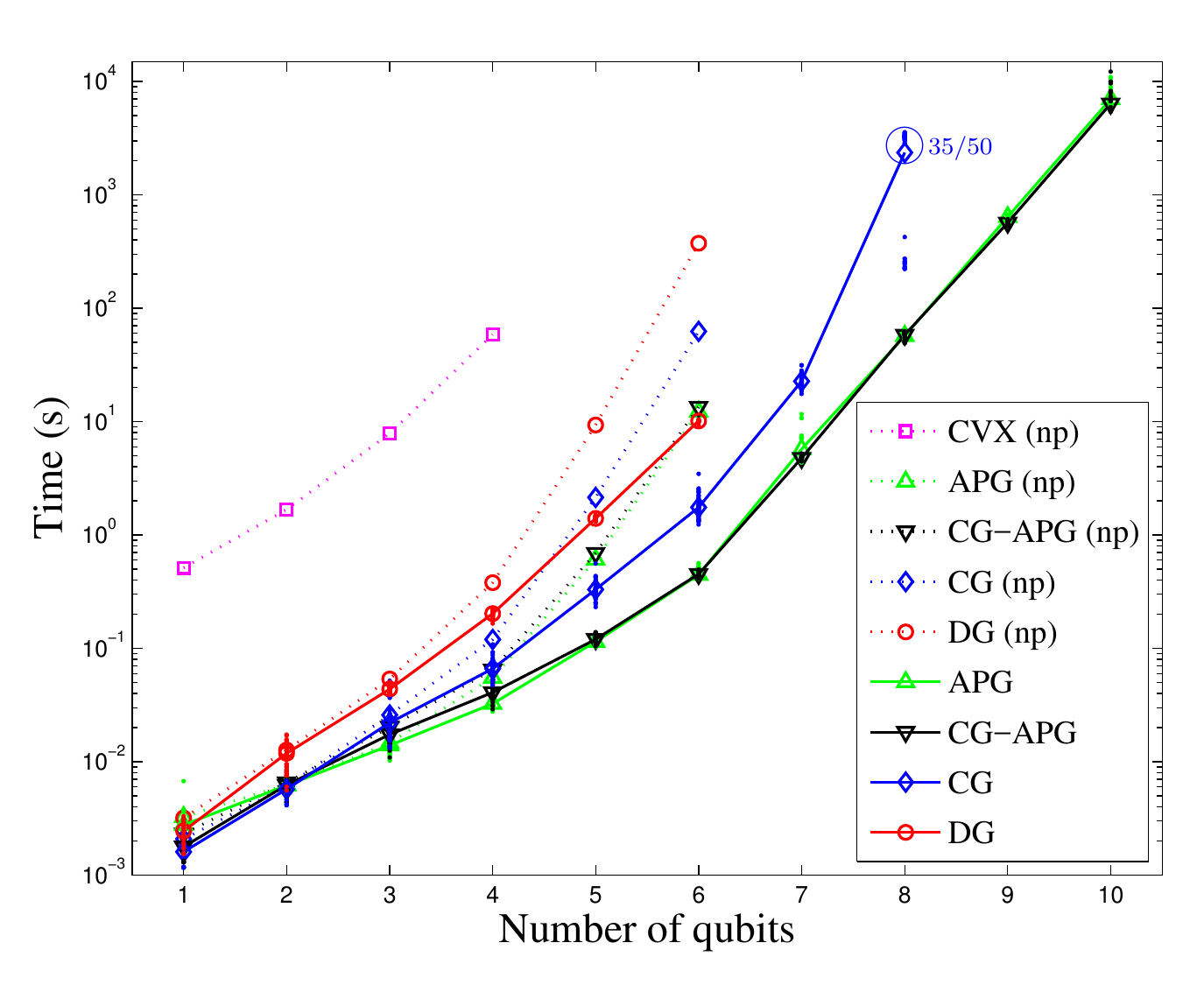}
\caption{\label{fig:nqb}%
Time taken, for a convergence criterion of $L/L_{\max} =99.9\%$, for different algorithms on a varying number of qubits $n$. For each $n$, 50 states are used, each a Haar-random pure state with 10\% white noise to emulate a noisy preparation. For each state, the algorithms are run for $\{f_k=p_k\}$, where $\{p_k\}$ are the Born probabilities for the state on the $n$-qubit product-Pauli POM. The MLE is hence the actual state.
The lines labeled ``np" indicate runs \emph{without} using the product structure. These stop at six qubits due to the long time taken. The lines are drawn through the average time taken for each algorithm over the 50 states; the scatter of the timings are shown only for the algorithms using the product structure. For $n=8$, CG did not converge within the maximum alloted time ($>\!\!20$ times that taken by CG-APG/APG) in 35 of the 50 states; these points (circled) are plotted with that maximum time, and the average time taken is a lower bound on the actual average for CG. CG failed to converge in a reasonable time for all states beyond 8 qubits; DG failed to converge beyond 6 qubits.
}
\end{figure}

Figure \ref{fig:nqb} shows the performance of the different algorithms for a varying number of qubits with and without exploiting the product structure, for the product-Pauli measurement. The APG and CG-APG algorithms show a substantial improvement in speed over other algorithms. With the faster speed, one can perform accurate MLE reconstruction for larger systems in the same amount of time. Observe that the CG-APG and APG runtimes are similar in Fig.~\ref{fig:nqb}, with CG-APG about 10\% faster than APG for $n=10$ (exploiting the product structure). This is because the advantage of CG-APG over APG occurs early on in the run, when the walk is far from the minimum. APG works well enough if one is seeking an accurate MLE so that most of the runtime is spent resolving the exact MLE location in the vicinity of the minimum. However, for long runtimes, the 10\% savings from CG-APG is not inconsequential. CG-APG combines the advantages of CG and APG, without much increase in implementation complexity, and works well even in cases like that of Fig.~1 where APG spends a long time wandering around far from the minimum.

Furthermore, significant speedup is visible when the product structure is incorporated. Exploiting the product structure is very different from putting in assumptions about the state or the noise: In the former, one knows the structure by design of the tomographic experiment; the latter requires checks of compliance, which need not be easy or even possible. Tomography experiments beyond a couple of qubits typically employ POMs with a product structure, because of the comparative ease in design and construction, so this product assumption often holds. 

For comparison, we display the runtime for the general-purpose CVX toolbox for convex optimization \cite{CVX1,CVX2}. The clear disadvantage there is the inability to capitalize on the product structure. All computations are conducted with {\sl MATLAB} on a desktop computer (3 GHz Intel Xeon CPU E5-1660).

The good performance of APG/CG-APG goes beyond the product-Pauli measurement. Appendix~\ref{AppD} gives results for the product-tetrahedron POM \cite{TetraPOM} and the symmetric, informationally complete (SIC) POM \cite{SICPOM1,SICPOM}, yielding similar conclusions. This is to be expected, as the slow convergence of DG/CG, remedied by the APG algorithm, is independent of the measurement choice.

\textit{Conclusion.}---
We have demonstrated that, with the right algorithm, MLE reconstruction can be done quickly and reliably, with no latent restriction on the accuracy of the MLE obtained, and no need for added, possibly unverifiable, assumptions. As the dimension increases, there is no getting around the fact that any tomographic reconstruction will become expensive, but our algorithm slows the onset of that point beyond the system size currently accessible in experiments. We note that our method can be immediately applied to process tomography. Furthermore, it is a general method for optimization in the quantum state space or other constrained spaces, and hence useful for such problems.

\textit{Remark:} Upon completion of our work, we came to be aware \cite{Leach} of Ref.~\cite{Goncalves}, an earlier work employing projected gradient techniques for optimization over the quantum state space. In particular, MLE reconstruction was investigated as an application. However, the discussion there was restricted to two- to four-qubit tomography, and the authors do not use accelerated gradients---as we do here---crucial for fast convergence, and certainly not our hybrid CG-APG method.

This work is funded by the Singapore Ministry of Education (partly through
the Academic Research Fund Tier 3 MOE2012-T3-1-009) and the National Research
Foundation of Singapore. The research is also supported by the National Research Foundation
(NRF), Prime Minister's Office, Singapore, under its CREATE programme,
Singapore-MIT Alliance for Research and Technology (SMART) BioSystems
and Micromechanics (BioSyM) IRG. HKN is partly funded by a Yale-NUS College start-up grant. We thank Christian Roos and Otfried G{\"u}hne for sharing the experimental data of Ref.~\cite{8qubit} and information about the MLE reconstruction used in that work. ZZ thanks Chenglong Bao for his discussions regarding APG and George Barbastathis for general discussions. We are also grateful to Andrew Scott, Michael Hoang, Christopher Fuchs and Markus Grassl for sharing the 7-qubit fiducial state for our SIC-POM example (Appendix~\ref{AppD}).

J.~Shang and Z.~Zhang contributed equally to this work.

\bigskip

\appendix
\section{Time taken for 8-qubit trajectories}\label{App0}

Here we show again the trajectories taken by different algorithms as in Fig.~1 in the main text, but now plotted against time rather than iterative steps.

\begin{figure}[h]
\includegraphics[trim=0mm 3mm 0mm 3mm, clip, width=\columnwidth,height=6.3cm]{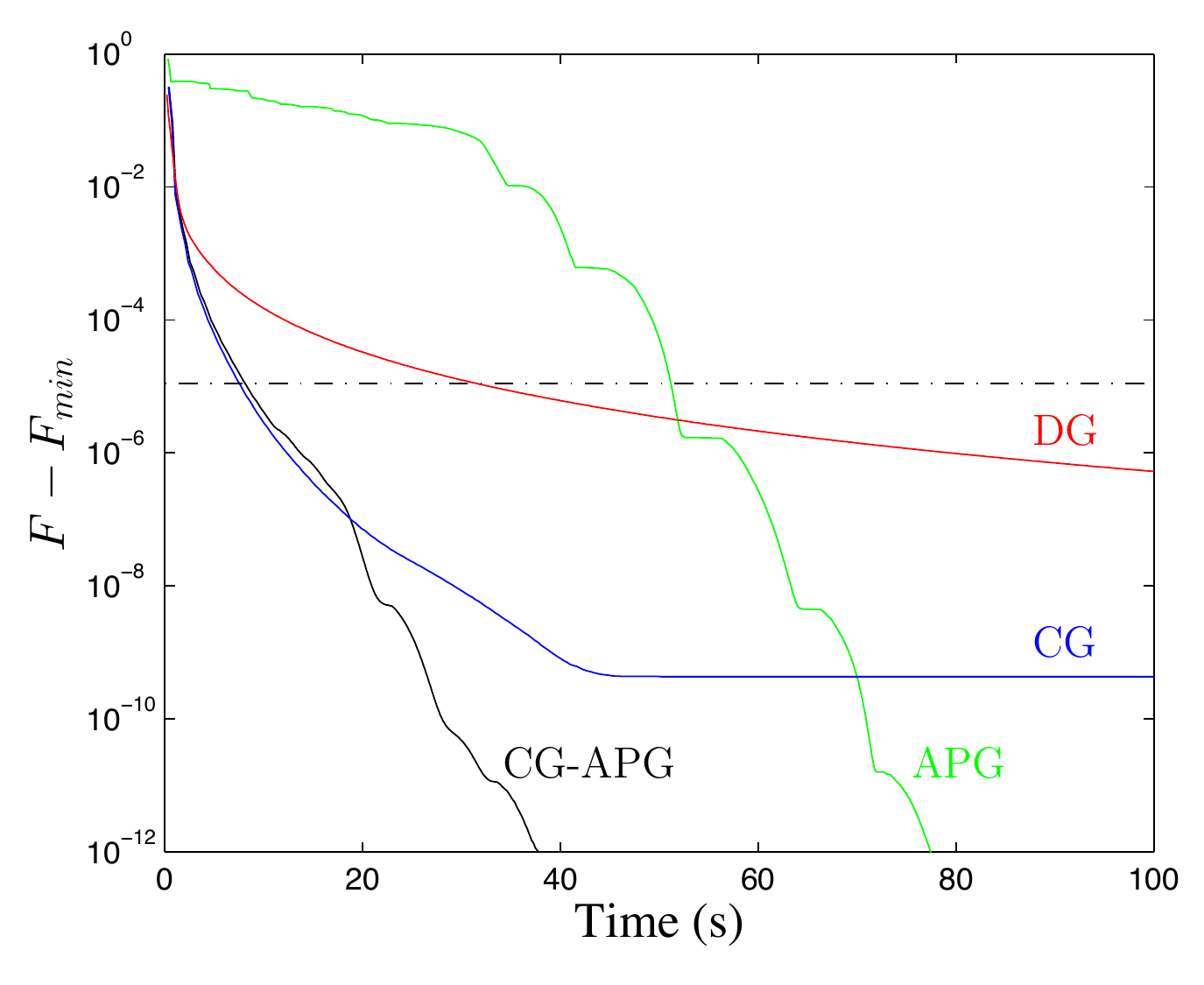}
\caption{\label{fig:SuppMat1}%
The deviation $F-F_{\min}=-\frac{1}{N}\log(L/L_{\max})$ versus time taken for different algorithms, for the experimental data of \cite{8qubit}.
}
\end{figure}

\section{The reconstructed 8-qubit state}\label{R8qb}
For the 8-qubit dataset, the 12 largest eigenvalues of the reconstructed MLE states from our CG-APG algorithm and the original H{\"a}ffner et al.~reconstruction using DG in the factored space \cite{8qubit} are given in Table \ref{tab:eig}.
\begin{table}[!h]
\caption{\label{tab:eig}The 12 largest eigenvalues of the reconstructed MLE states.}
\begin{tabular}{c@{\hspace*{0.8cm}}c}
Our reconstruction & H{\"a}ffner et al.~\\
(using CG-APG)&(Ref.~\cite{8qubit}, using DG)\\
\hline
0.7512&0.7514\\
0.0609&0.0609\\
0.0458&0.0456\\
0.0403&0.0400\\
0.0237&0.0234\\
0.0193&0.0189\\
0.0178&0.0174\\
0.0153&0.0149\\
0.0106&0.0102\\
0.0051&0.0048\\
0.0039&0.0036\\
0.0030&0.0030
\end{tabular}
\end{table}
Observe the close correspondence between the two reconstructed states. 

The table of eigenvalues also demonstrate the problem with using the compressed sensing (CS) scheme of Ref.~\cite{CompSens}. The CS approach requires an \emph{a priori} choice in the rank of the reconstructed state; specifically, it works well when that choice is one of low rank. Looking at the list of eigenvalues above, we see that one either uses a rank-1 state, or one should include many more eigenvalues, as the subsequent ones are comparable in size. Without access to an unrestricted reconstruction, attainable only by our CG-APG or other full MLE schemes, there is no way of making a verifiable rank choice for the CS approach.

\section{The APG and CG-APG algorithms}\label{AppA}
We discuss various technical details pertaining to the APG and/or CG-APG algorithms described in the main text.

\subsection{The projection algorithm used in APG}
As explained in the main text, the APG algorithm relies on a projection $\proj(\cdot)$ to enforce the quantum constraints after each gradient step. The argument of $\proj(\cdot)$ is a Hermitian operator $\varrho$ with eigenvalues $\lambda_i$ (in descending order) and eigenvectors $|\psi_i\rangle$. One projects $\{\lambda_i\}$ onto the probability simplex so that ${\{\lambda_i\}\rightarrow\{\overline\lambda_i\}}$ with ${\overline\lambda_i\geq 0}\,\,\forall i$ and ${\sum_i\overline\lambda_i=1}$, and then rebuilds the operator with $\{\overline\lambda_i\}$, i.e., ${\proj(\varrho)=\sum_i\overline\lambda_i|\psi_i\rangle\langle\psi_i|}$.
The projection of $\lambda_i$ onto the simplex is done as follows \cite{GaussianNoise}:
Find ${u=\max\left\{j:\lambda_j-\frac1{j}\bigl(\sum_{i=1}^j\lambda_i-1\bigr)>0\right\}}$, then define
${w=\frac1{u}\left(\sum_{i=1}^u \lambda_i-1\right)}$. Finally we have $\overline\lambda_i=\max\{\lambda_i-w,0\}$.

\subsection{Handling negative $p_k$ values}

During the gradient step of APG, one can wind up outside the physical state space, i.e., $\varrho_i$ at each iterative step need not be a valid state. It can even happen that not all $p_{k,i}\equiv \tr(\varrho_i\Pi_k)$s needed in the iterative step are positive, for which $F(\varrho_i)$ is ill-defined because of the logarithm. We can prevent this by checking whether any $p_{k,i}$ is negative after $\varrho_i$ is computed, and set $\varrho_i=\rho_i$ if this happens to be the case. Empirically, we observe such cases to occur only very rarely.

\subsection{Convergence tweaks for APG}\label{SuppMat:opt}

We also incorporated a few small adjustments to APG recommended in
Ref.~\cite{TFOCS}, as well as the Barzilai-Borwein (BB) method for
computing step sizes \cite{BBStep}, for better step-size estimation and improved performance in the implementation of the CG-APG algorithm
used to produce the figures in the main text. We list those adjustments here.

First, for iterative step $i>1$, rather than fixing the step size as $t_i=t_{i-1}$, we set \cite{BBStep}
\begin{equation}
t_i = \frac{\langle \varrho_i-\varrho_{i-1},\nabla F(\varrho_i) - \nabla F(\varrho_{i-1})\rangle}{\langle \nabla F(\varrho_i) - \nabla F(\varrho_{i-1}),\nabla F(\varrho_i) - \nabla F(\varrho_{i-1})\rangle}
\end{equation}
if there was no restart in the previous iteration and the denominator
is nonzero; otherwise we set $t_i=\alpha t_{i-1}$, for some pre-chosen constant $\alpha$. We used $\alpha=1.1$ and $\beta=0.5$ (see main text) as recommended in \cite{TFOCS}.

We also use the following update on $\theta_i$ and $\varrho_i$ for $i>1$ to prevent
changes in $t_i$ from affecting convergence:
\begin{subequations}
\begin{eqnarray}
\hat\theta_{i-1} &=& \theta_{i-1}\sqrt{t_{i-1}/t_i}\\
\label{5b}\theta_i &=& \frac{1}{2}{\left(1+\sqrt{1+4\hat\theta_{i-1}^2}\right)}\\
\label{5c}\varrho_i&=& \rho_i+\hat\delta_i(\hat\theta_{i-1}-1)/\theta_i
\end{eqnarray}
\end{subequations}
The rules Eqs.~\eqref{5b} and \eqref{5c} are exactly those stated in the APG algorithm in the main text, but with $\theta_{i-1}$ replaced by $\hat\theta_{i-1}$ for the step-size adjustment.

Sometimes we observe that standard APG as prescribed by \cite{AdaptRes} fails to restart early enough for good performance. We hence use a stricter restart criterion: Restart when
\begin{align}
\frac{\tr(\delta_i\hat{\delta_i})}{\sqrt{\tr(\delta_i^2)\tr\bigl(\hat\delta_i^2\bigr)}} < \gamma,
\end{align}
with $\gamma$ set to a small positive value ($0.01$ for the graphs in
the main text).

We notice that the BB method can sometimes give a larger variation in performance of the APG algorithm for different data. This is visible in Fig.~2 of the main Letter, where the 7-qubit situation for APG shows a slightly larger scatter than for other $n$ values. The scatter is reduced when the restart parameter $\gamma$ is set larger so that the restart occurs earlier, suggesting a possible interference with adaptive restart. One can thus adjust the $\gamma$ parameter if one is concerned about the variation in performance; or one can turn off the BB step-size optimization altogether, and see only a very slight worsening of the average runtime; see Fig.~\ref{fig:SuppMatBB} below. Note that we did not see this variation in performance for CG-APG with the BB method for all cases tested.

\begin{figure}[!h]
\includegraphics[trim=3mm 0mm 5mm 5mm, clip, width=\columnwidth]{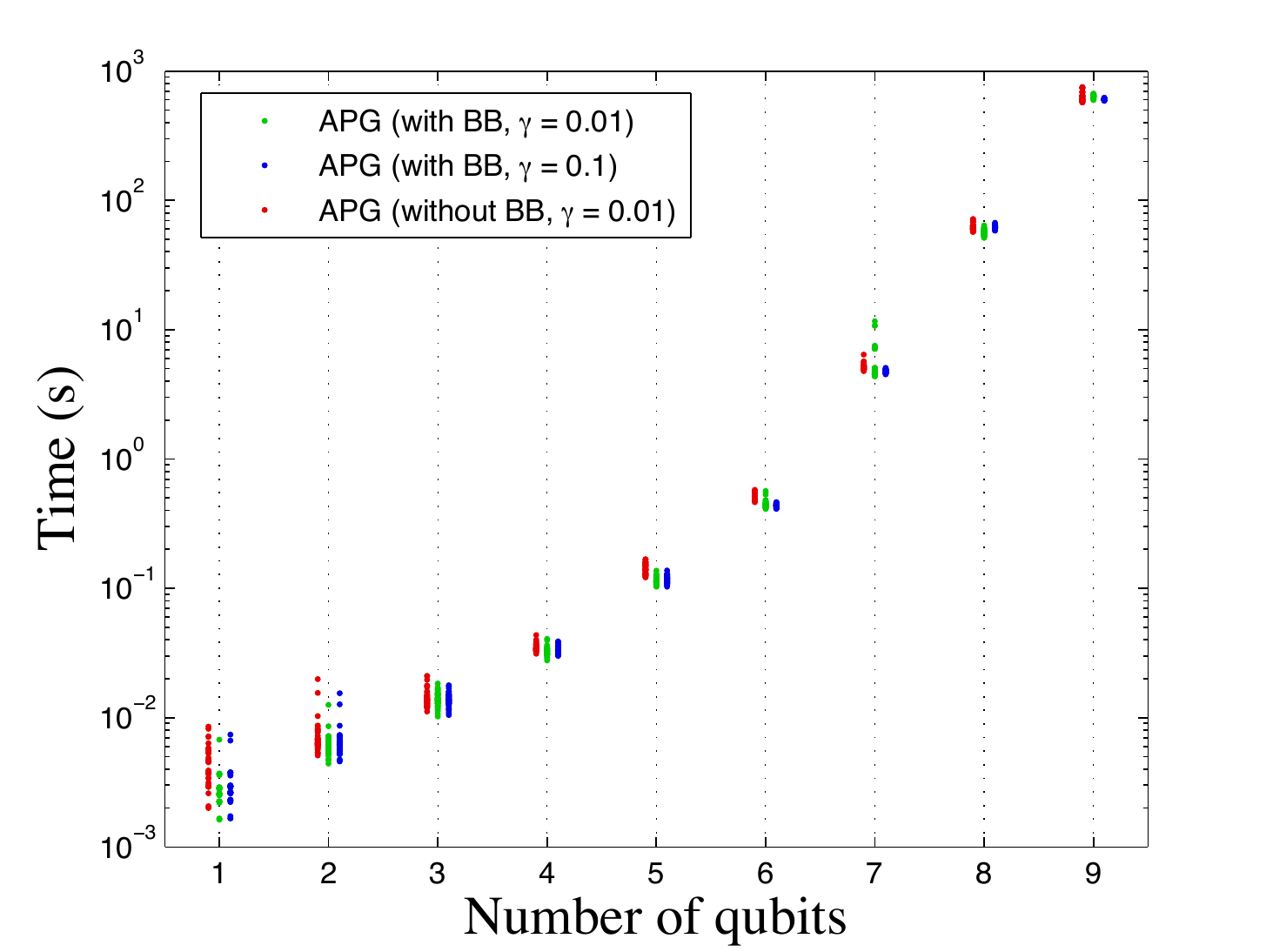}
\caption{\label{fig:SuppMatBB}%
The performance of the APG algorithm with and without the BB step-size optimization, and with different values of $\gamma$. The data points for the three cases for the same $n$ are plotted with a slight horizontal offset for better visibility. Observe, in particular, the scatter in the runtimes for $n=7$ and $\gamma=0.01$ when the BB method is used.
}
\end{figure}

\subsection{Switchover criterion for CG-APG}
For the CG-APG algorithm, as explained in the main text, one would like to start with CG iterations and switch to APG when the Hessian stabilizes, i.e., it changes only by a little with further APG steps. This happens when the trajectory is sufficiently close to the MLE. Here, we explain the technical details of this switchover.

The Hessian of $F(\rho)$---its curvature---characterizes its local quadratic structure. It is the ``second derivative" of $F(\rho)$, and comes from considering the second-order variation of $F$: $\delta^2F(\rho)\equiv \delta F\bigl(\rho+\widetilde{\delta\rho}\bigr)-\delta F(\rho)$, where $\delta F(\rho)\equiv -\tr\Bigl(\delta\rho\, R(\rho)\Bigr)$, the first-order variation of $F$, with $R(\rho)=\sum_k\Pi_kf_k/p_k$ and $p_k\equiv \tr(\rho\Pi_k)$ as in the main text. Here, $\delta\rho$ and $\widetilde{\delta\rho}$ are independent infinitesimal variations of $\rho$. A little algebra gives
\begin{equation}
\delta^2F(\rho)=\tr{\left(\delta\rho\sum_k\Pi_k\tr\Bigl(\widetilde{\delta\rho}\,\Pi_k\Bigr)\frac{f_k}{p_k^2}\right)},
\end{equation}
and we identify the linear operator [on $\cB(\cH)$]
\begin{equation}
H(\rho;\,\cdot\,) = \sum_k \Pi_k \tr(\,\cdot\,\Pi_k )\frac{f_k}{p_k^2}
\end{equation}
as the Hessian of $F$ at $\rho$.

The eigenvalues of $H$ give the local quadratic structure of $F(\rho)$. Ideally,
determining the right time during CG to switch to APG requires computing how much $H$ changes across successive APG steps from the current value of $\rho$. However, this would be very costly: It is as if one is running APG alongside CG, and the Hessian is a large matrix ($d^2\times d^2$ in size) and hence expensive to compute.

Instead, we adopt a compromise that works well in practice: (1) we treat
the $\Pi_k$s as if they were all mutually orthogonal so that the eigenvalues of
$H$ would be equal to $\{f_k/p_k^2\}$, and (2) we
look at the change in $H$ between iterations of CG instead of
between iterations of APG.
The $\Pi_k$s are never exactly mutually orthogonal for informationally complete measurements, but a good tomographic design would seek to spread out the $\Pi_k$ directions, and for large dimensional situations, their mutual overlaps will be small and $\{f_k/p_k^2\}$ is a good enough proxy for the eigenvalues of the
Hessian.
While looking at the change in $H$ across iterations of CG would not always guarantee a similar change for APG, a small change with CG iterations signals closeness to the MLE, or that CG has stagnated. In either case, one should switch to APG.

Thus, in our implementation of CG-APG, we first initialize CG with the
maximally mixed state, and switch to APG at the first iteration when the overlap,
\begin{equation}
\frac{\vec q_i\cdot\vec q_{i-1}}{\sqrt{|\vec q_i|^2|\vec q_{i-1}|^2}},
\end{equation}
exceeds $\cos\phi$ for some chosen small $\phi$ value. Here, $\vec q_i\equiv (f_1/p_{1,i}^2,f_2/p_{2,i}^2,\ldots,f_K/p_{K,i}^2)$, where $p_{k,i}=\tr(\rho_i\Pi_k)$ for state $\rho_i$ of the $i$th CG iteration. The switchover thus occurs when the angle between the $\vec q$s for subsequent iterations is small enough.
We find that $\phi=0.01$ radians works well in practice.

\section{Trajectories for generic states}\label{AppB}

\begin{figure}[h]
\includegraphics[trim=0mm 3mm 0mm 3mm, clip, width=\columnwidth,height=6.3cm]{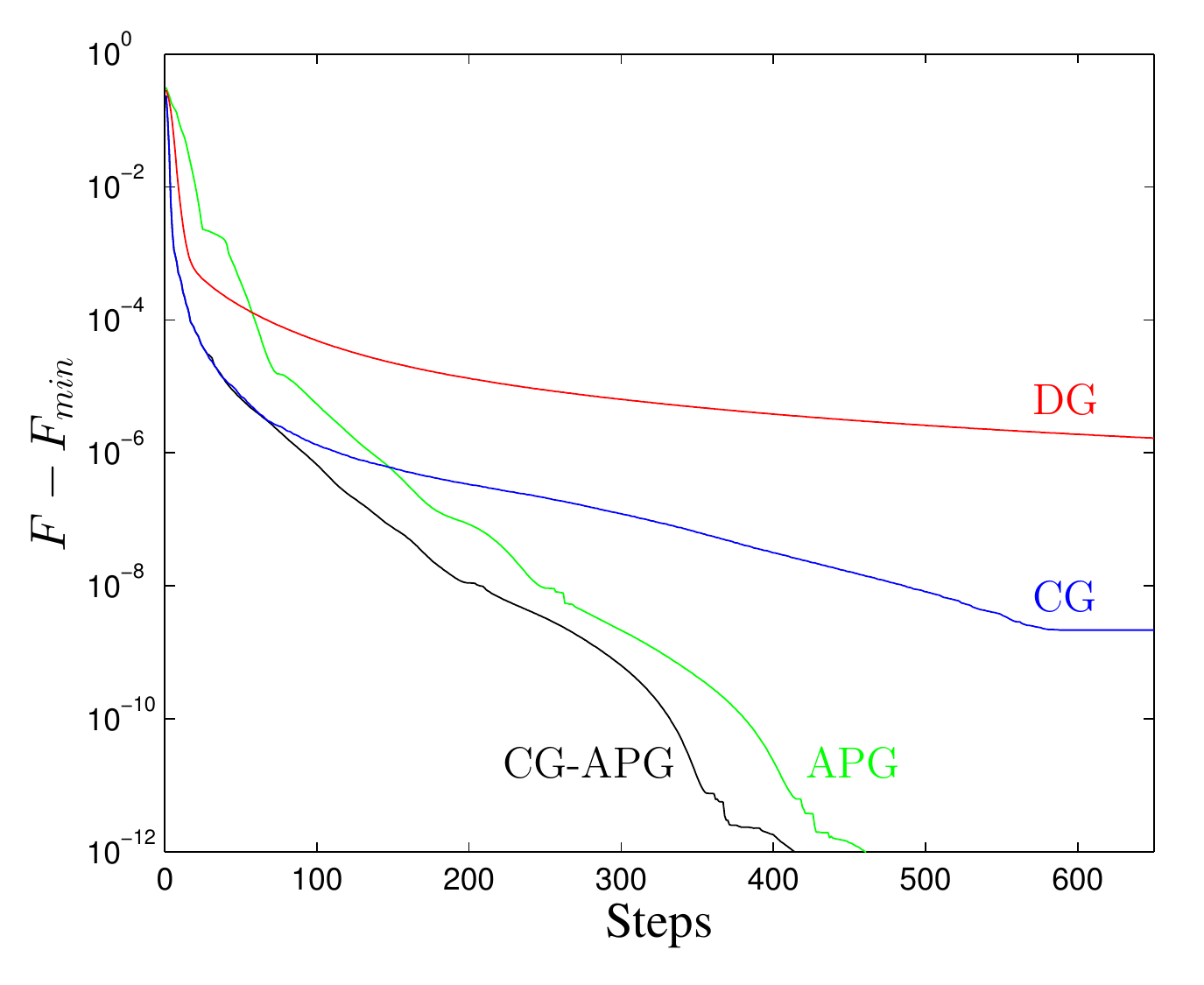}
\caption{\label{fig:SuppMat2}%
The plot shows the trajectories taken by different algorithms as in Fig.~1 in the main text, but for simulated data generated from a random 8-qubit pure state with 10\% added white noise. }
\end{figure}

Figure 1 in the main text shows the trajectories taken by different algorithms for the experimental data of Ref.~\cite{8qubit} for a noisy $W$-state. There, we saw a long initial slow phase of APG, which is in fact atypical of the behavior seen for generic states. Figure \ref{fig:SuppMat2} shows the more representative behavior for a random 8-qubit pure state with 10\% added white noise.  As in Fig.~1, $N=3^8\times 100$, with 100 copies for each of the $3^8$ settings of the 8-qubit product-Pauli POM. Observe the significantly shorter length of the initial slow phase of APG than for the noisy $W$-state in Fig.~1.

\section{Exploiting the product structure: Computational savings}\label{AppC}
Here, we present the counting argument that gives $O(K_r^{n+1})$ as the computational cost of evaluating a full set of Born probabilities after making use of the product structure of the POM. To remind the reader of the notation: The system comprises $n$ registers each of dimension $d_r$; the POM on each register is $\{\pi_k\}_{k=1}^{K_r}$; the $n$-register POM outcome is $\Pi_{\vec k}=\pi_{k_1}\otimes\pi_{k_2}\otimes\ldots\otimes\pi_{k_n}$, with $\vec k\equiv (k_1,k_2,\ldots ,k_n)$ and $k_a=1,\ldots,K_r$; and $\rho_{\ell}^{(k_{\ell+1},\ldots,k_n)}\equiv \mathrm{tr}_{\ell+1}\{\ldots\mathrm{tr}_{n-1}\{\mathrm{tr}_n\{\rho\pi_{k_n}\}\pi_{k_{n-1}}\}\ldots\pi_{k_{\ell+1}}\}$. We also need the following basic fact: Evaluating $\tr\{AB\}$ for $A$ an $n\times m$ matrix and $B$ an $m\times n$ matrix requires $2mn$ operations (elementary addition/multiplication).

In each step of the procedure described in the main text, one needs to evaluate $\rho_{\ell-1}^{(k_{\ell},\ldots,k_n)}=\tr_{\ell}\{\rho_{\ell}^{(k_{\ell+1},\ldots,k_n)}\pi_{k_\ell}\}$ for given $k_\ell,\ldots,k_n$. One such evaluation requires the computation of the trace of $\pi_{k_\ell}$ with each of the $(d_r)^{\ell-1}\times (d_r)^{\ell-1}$ submatrices of $\rho_{\ell}^{(k_{\ell+1},\ldots,k_n)}$. $\pi_{k_\ell}$ and each submatrix are $d_r\times d_r$ in size, so the computational cost of evaluating $\rho_{\ell-1}^{(k_{\ell},\ldots,k_n)}$ is $2d_r^2\times d_r^{2(\ell-1)}=2d_r^{2\ell}$ operations. One incurs this cost for every choice of $k_\ell,\ldots,k_n$, so the total cost of evaluating $\rho_{\ell-1}^{(k_{\ell},\ldots,k_n)}$ for all $k_\ell,\ldots,k_n$, for given $\ell$, is $2K_r^{n-\ell+1}d_r^{2\ell}$. Adding up this cost over all values of $\ell=1,2,\ldots,n$ gives the total cost for evaluating a full set of Born probabilities as
\begin{eqnarray}
2\sum_{\ell=1}^n K_r^{n-\ell+1}d_r^{2\ell}&=&2K_r^nd_r^2\sum_{\ell=0}^{n-1}{\left(\frac{d_r^2}{K_r}\right)}^\ell\nonumber\\
&=&2K_r^nd_r^2{\left[\frac{1-(d_r^2/K_r)^n}{1-(d_r^2/K_r)}\right]}.
\end{eqnarray}
For $K_r> d_r^2$, as is usually the case, this gives the dominant computational cost of $O(K_r^{n+1})$; for $K_r=d_r^2$, one has instead the cost of $O(nK_r^{n+1})$.

\section{Performance for different measurements}\label{AppD}

\begin{figure}[!h]
\includegraphics[trim=3.5mm 3.5mm 4mm 0mm, clip, width=\columnwidth]{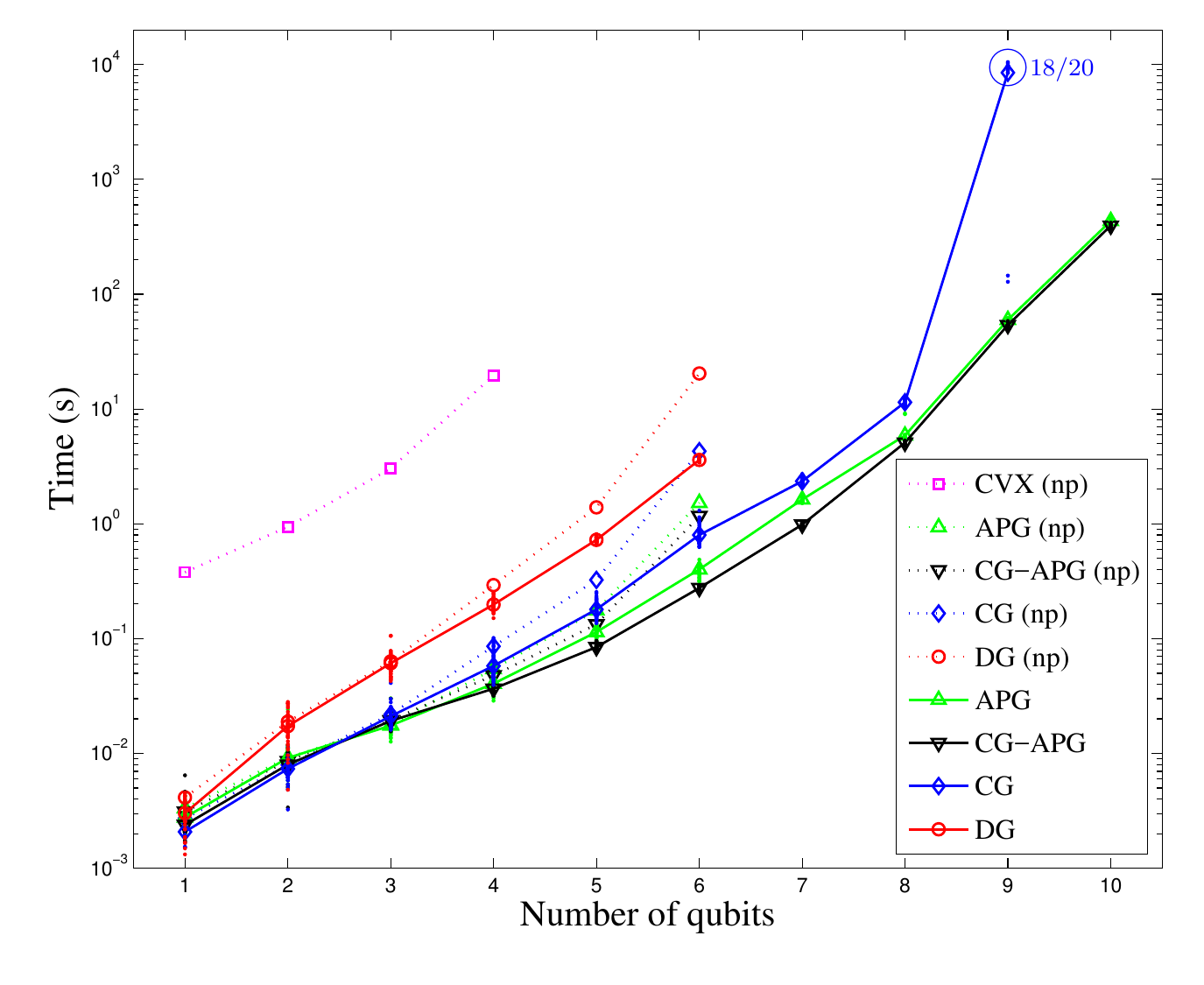}
\caption{\label{fig:tetra}%
Time taken for different algorithms on a varying number of qubits $n$ for the product-tetrahedron POM. For the $n=9$ case of the CG algorithm, due to the long runtime, only 20 trial states were used. CG did not converge within the maximum alloted time for 18 of those 20 states. CG failed to converge in a reasonable time for all states for 10 qubits; DG failed to converge beyond 6 qubits.}
\end{figure}

\begin{figure}[!h]
\includegraphics[trim=8mm 3.5mm 15mm 10mm, clip, width=\columnwidth]{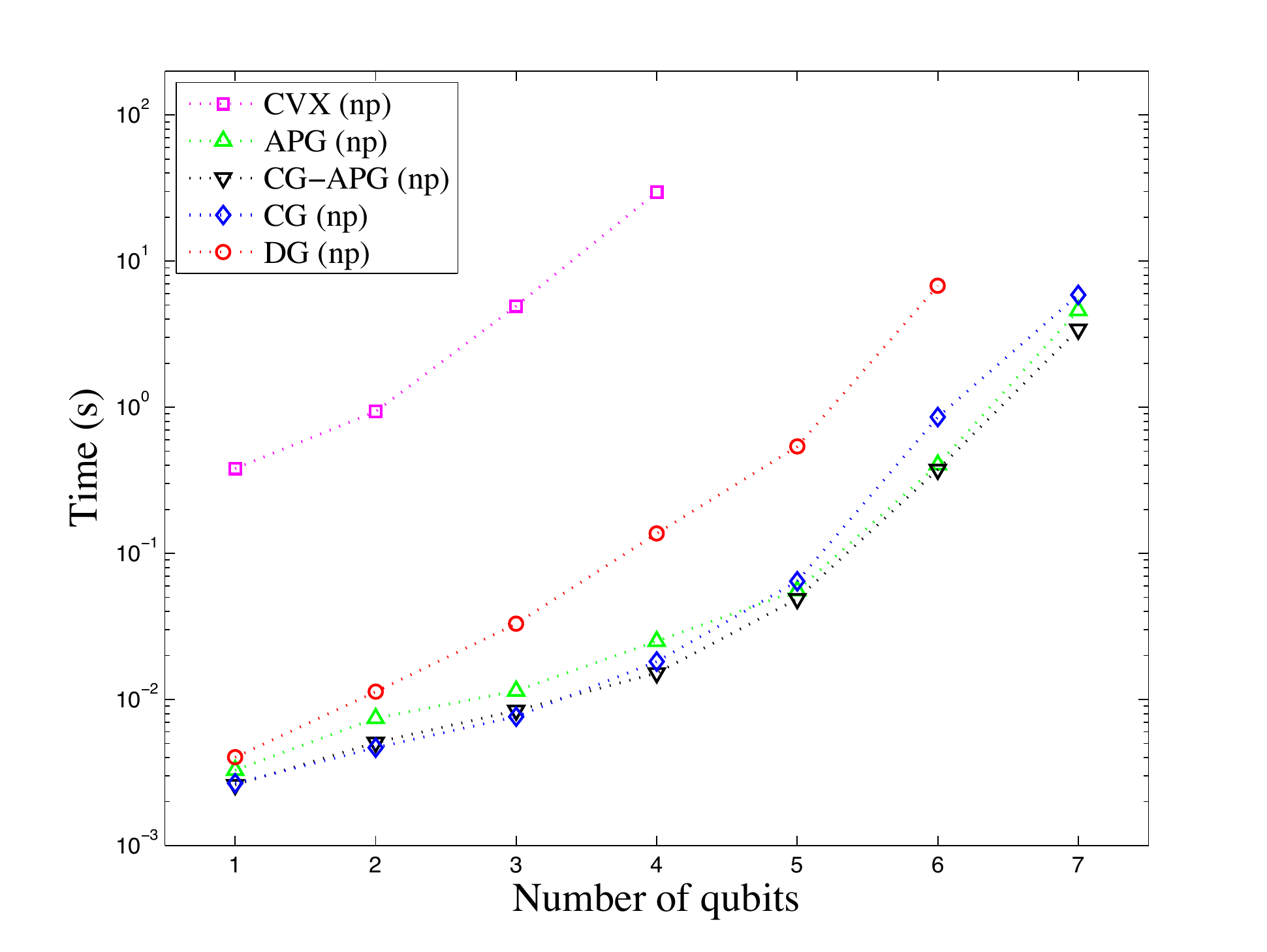}
\caption{\label{fig:SIC}%
Time taken for different algorithms on a varying number of qubits $n$ for the SIC-POM. As the SIC-POMs do not possess any product structure, only the ``np'' lines are applicable. The simulations stop at 7 qubits as we do not have the 8-qubit (and beyond) fiducial state needed for the SIC-POM construction.}
\end{figure}

As stated in the main text, the edge our APG and CG-APG algorithms has over the DG and CG algorithms (in the factored space) does not depend on the measurement choice. Here, we demonstrate this point by benchmarking the performance of APG and CG-APG against that of DG and CG for two more measurements (in addition to the product-Pauli POM in the main text): the product-tetrahedron POM \cite{TetraPOM} in Fig.~\ref{fig:tetra}, and the symmetric, informationally complete POM (SIC-POM) \cite{SICPOM1,SICPOM} in Fig.~\ref{fig:SIC}.

The parameters in both figures used are identical (with the exception of $n=9$ of the product-tetrahedron POM for CG) to those of Fig.~2 in the main text, repeated here for the reader's convenience: 
For each $n$, 50 states are used, each a Haar-random pure state with 10\% white noise to emulate a noisy preparation. For each state, the algorithms are run for $\{f_k=p_k\}$, where $\{p_k\}$ are the Born probabilities for the state on the respective $n$-qubit POM. The MLE is hence the actual state. The lines labeled ``np" indicate runs \emph{without} using the product structure. The lines are drawn through the average time taken for each algorithm over the 50 states; the scatter of the timings are shown only for the algorithms using the product structure.\\

We note that the performance of the CG algorithm for the SIC-POM showed exceptional sensitivity, not observed in other cases, to the parameters that enter the line minimization of the CG algorithm. The CG plot of Fig.~\ref{fig:SIC} is given for line-minimization parameters optimized for the best CG runtime. While the previous parameters used for the product-Pauli measurement worked also for the product-tetrahedron POM (in fact, there we see little difference in the performance for different parameter choices), the CG runs did not converge at all unless we tweak the parametrs. We do not yet understand the underlying reason for this sensitivity, and it may be mitigated with the use of a more elaborate line-minimization procedure, but one should perhaps take this as an added note of caution when using the CG algorithm.


\begin{thebibliography}{00}
\bibitem{Leach}
We are in gratitude to J.~Leach for alerting us to Ref.~\cite{Goncalves}.
\end{thebibliography}
\end{document}